\documentclass[prx,reprint,superscriptaddress,twocolumn,showkeys]{revtex4-1}
\usepackage[utf8]{inputenc} 
\usepackage{amsmath}
\usepackage{braket}
\usepackage{graphicx}
\usepackage{pgfplots}
\pgfplotsset{compat=1.15}
\usepackage{csquotes}
\usepackage{hhline}
\usepackage{amssymb}

\usepackage{dcolumn}
\usepackage{tabularx}
\usepackage[colorlinks=true,linkcolor=blue,citecolor=blue,urlcolor=blue]{hyperref}
\usepackage{braket}
\usepackage{float}
\usepackage{listings}
\usepackage{color}
\usepackage{subcaption}
\usepackage{placeins}
\definecolor{mygreen}{rgb}{0,0.6,0}
\definecolor{mygray}{rgb}{0.5,0.5,0.5}
\definecolor{mymauve}{rgb}{0.58,0,0.82}

\newcolumntype{C}{>{\centering\arraybackslash}X}

\begin{document}
\title{Experimental Realization of Quantum Darwinism State on Quantum Computers}

\author{Rakesh Saini}
\email{rs474390@gmail.com}
\affiliation{Department of Physics, \\ Indian Institute of Technology, (ISM), Dhanbad 826004, Jharkhand, India}
\author{Bikash K. Behera $^{a}$}
\email{bikash@bikashsquantum.com}
\thanks{$^{a}$Corresponding author}
\affiliation{Bikash's Quantum (OPC) Pvt. Ltd., Balindi, Mohanpur 741246, Nadia, West Bengal, India}

\begin{abstract}
It is well-known that decoherence is a crucial barrier in realizing various quantum information processing tasks; on the other hand, it plays a pivotal role in explaining how a quantum system's fragile state leads to the robust classical state. Zurek [Nat. Phys. \textbf{5}, 181-188 (2009)] has developed the theory which successfully describes the emergence of classical objectivity of quantum system via decoherence, introduced by the environment. Here, we consider two systems for a model universe, in which the first system shows a random quantum state, and the other represents the environment. We take 2-, 3-, 4-, 5- and 6-qubit quantum circuits, where the system consists of one qubit and the rest qubits represent the environment qubits. We experimentally realize the Darwinism state constructed by this system's ensemble on two real devices, ibmq\_athens and ibmq\_16\_melbourne. We then use the results to investigate quantum-classical correlation and the mutual information present between the system and the environment.
\end{abstract}

\maketitle

\section{Introduction}
Quantum mechanics is a fascinating but weird theory, which is inconsistent with the classical one, and its properties show its weirdness \cite{qd_SchrodingerNaturwissenschaften1935,qd_Wolchover2018}. One of the fundamental properties is superposition, which says that instead of localizing in space (as we see usually), the particle exists at different places/states simultaneously with some probability. As an example, we can consider the cat Dead and Alive model proposed by Schr$\ddot{o}$dinger \cite{qd_SchrodingerNaturwissenschaften1935}. This superposition property governs the system and always exists in the quantum realm. The superposition states are highly fragile, and due to their fragile nature, if we interact with the quantum system, we accidentally destroy its coherence. As a result, it loses its peculiarity and collapses in one of the probable states (pointer state/survival state) \cite{qd_Zureknature2009}. This interaction is successfully understood by the decoherence theory \cite{qd_JoosSpringer2003,qd_ZurekRMP2003,qd_SchlosshauerSpringer2007}. For example, when we bombard electrons on to slits, they interfere with each other and show interference on the screen, however, if we decohere the system by observing the path of the electrons, the interference vanishes, and the observed pattern is the same as in the classical objects \cite{qd_Wolchover2018,qd_physicsworld}.

The decoherence theory plays a vital role in explaining the rise of classical objectivity of the quantum system \cite{qd_ZurekRMP2003,qd_KorbiczPRL2014}. The quantum systems are very hard to keep isolate from the surrounding environment. Because there is always a leak of information from the system to the environment. Furthermore, the quantum system loses its peculiarity or decohered during this leak. During this type of leak or interaction, the information about the system imprints on the environment. Thus, via accessing the environment, we can obtain information about the system without interacting with it. However, the decoherence theory does not contain this type of explanation about the environment. It merely traces it out and treats this information as inaccessible and irrelevant. Zurek was the first to successfully explain these types of cases and gave a theory named the quantum Darwinism theory \cite{qd_Zureknature2009,qd_RiedelPRL2010}. Quantum Darwinism explains why we observe the world as classical/robust instead of fragile (as in the quantum realm). It explains that most of the information we extract about the system comes from the environment. Furthermore, it uses the environment as the communication channel \cite{qd_Zureknature2009,qd_KohoutPRA2006,qd_ZurekPTRSA2018} between the system and the observers.

This access of information via the environment comes from the entanglement concept, as we know that when two quantum particles are entangled, they carry the information about each other. Moreover, via measuring one particle, we can predict the information about the other one \cite{qd_wiki}. Following this concept, we can understand that entanglement is generated via interaction between the system and the environment, and the degree of entanglement depends upon the strength of the interaction between the system and the environment. Hence, consecutively, the information about both the system is interdependent after the interaction. Furthermore, measuring one of them can predict information about the rest.

IBM quantum experience has become a world-wide easily accessible quantum computing research platform, where experiments are carried out on real quantum computers consisting of 1 qubit, 5 qubits, 7 qubits, and 15 qubits. Among them 5-qubits and 15-qubits real chips are mostly used by the researchers to perform their quantum computational and quantum informational tasks. A python module named QISKit \cite{qd_qiskit} is also introduced to write programs, for running the quantum circuits on real devices and to collect the results. A larger number of tasks have already been performed such as quantum simulation \cite{qd_NoriNpjQI2020,qd_KapilarXiv2018,qd_ManabputraQIP2020}, quantum machine learning \cite{qd_hunaturescience2018,qd_MishraAISC2020,qd_KishoreQIP2020}, quantum algorithms \cite{qd_wei2018,qd_mandviwallaieee2018,qd_DuttaIETQC2020}, quantum communication \cite{qd_behraspringer2019,qd_BeheraQIP2019,qd_BarikIETQC2020}, quantum cryptography \cite{qd_saini2019,qd_WarkeQIP2020}, quantum circuit architecture \cite{qd_willeieee2019,qd_azad2019,qd_zulehnerieee2018}, quantum game \cite{qd_anandspringer2020, qd_palepi2020} to name a few.

Recently, on a photonic quantum simulator \cite{qd_LloydScience1996,qd_GuzikNP2012} the quantum Darwinism state has been successfully tested by Chen \emph{et al.} \cite{qd_Chen2019}. They have considered a six photonic system in their work in which one photon is used as the system, and the rest is for the environment. They have discussed the obtained quantum and classical information separately and mutually for two different angle parameters. In our article, we start by briefly discussing the Darwinism theory and its state. Then we design a quantum circuit on the IBM quantum experience platform using the real devices, a 5-qubit one ``ibmq\_athens" and a 15-qubit one ``ibmq\_16\_melbourne" to produce the quantum Darwinism state, and execute them for two different interaction strengths A and B. We have considered two-, three-, four-, five-, and six-qubit systems for realizing the Darwinism state, where in each case one qubit is used for the system qubit and the rest are the environment qubits. In each qubit system, we have investigated the behavior of the system-environment by considering different numbers of environmental fragments separately and together. Then we have compared the behavior induced in one system-rest environment and one system- rest fragments of the environment. Furthermore, we have performed the quantum state tomography \cite{qd_Altepeter2004} technique for each of the qubit systems and calculated mutual information (MI), quantum-correlation (discord), and classical correlation (Holevo quantity) presented between the system and the environment. All the above quantities are calculated and presented after applying the mitigation process to improve our executed results.

The rest of the paper is organized as follows. In section \ref{qd_Sec2}, we briefly discuss the quantum Darwinism theory and its state; in section \ref{qd_Sec3}, we present the circuit formation for quantum Darwinism state with the execution on real quantum devices available on IBM quantum experience platform. In section \ref{qd_Sec4}, we describe the mutual, quantum, and classical information theory, and calculate the quantities from the experimental results, then compare among different qubit systems.

\section{Quantum Darwinism\label{qd_Sec2}}
It is well known based on scientific experiments that the fundamental and extraordinarily tiny particles build the whole universe. These particles are on the quantum scale and show quantum properties (like- superposition). However, we can not observe these quantum properties in the physical world; we see the whole world as classical (violating quantum rules). So where this all quantumness go? Why can not we see it? Why do we see the state/position robust instead of fragile, as we see in the quantum? The detailed explanation of these questions lies in quantum Darwinism theory.

\begin{figure}[]
\includegraphics[scale=0.3]{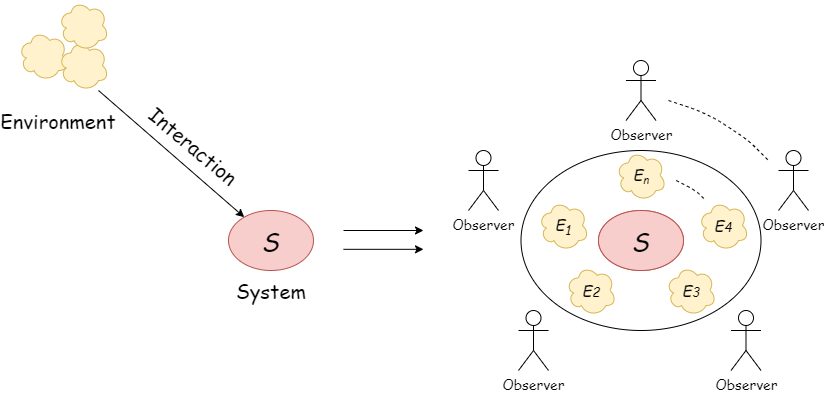}
\caption{The figure shows the interaction between the system and the environment. This interaction forms an ensemble where the system is in a sea of environment. This ensemble is then observed by the independent/with no prior consensus observers who can only have access to the environment. After the observers agree on the outcomes of these observations, classical objectivity arises.}
\label{qd_Fig1}
\end{figure}

Fig. \ref{qd_Fig1} shows a system, environment, and different independent observers (independent means there is no pre-agreement between the observers regarding measurement). There is a well-suitable condition for the environmental particles, which is non-interacting because if they interact with each other, then the stored information of the system will be mixed and become harder to measure correctly. So, when the non-interacting particles of the environment interact with the system, they work like a blank sheet. The system imprints the information of various possible states on them.

From the above discussion, the whole particles (which interact with the system) are the witness to the system's state, and observing this particle can re-prepare the system's state. To re-prepare the system's state, the system should imprint multiple redundant copies of information on the environmental particles, which is essential for the objectivity \cite{qd_ZwolakPRA2010}. Because observers can not observe the whole environment, only a tiny fragment of the environment is accessible. The observers can simultaneously measure the different environmental fragments from various positions and agree on the outcome information. As a result, the system's objectivity arises, and objectivity indicates the classicality. The states that come forward in this observation procedure, known as pointer states. Pointer states are the selected stable states by the system's environment that successfully imprints several redundant copies in respect to other possible states on the environment and the states that survive during the interaction (means did not get mixed in the environment) with the environment till they observed.

Now, the question arises that how much information does an environmental fragment contain about the system? Or what size of the fragment have all of the information about the system? These questions can be answered by calculating the mutual information between the system and the environmental fragments.

\begin{eqnarray}
I(S:F) = H_S + H_F - H_{S,F}
\label{qd_Eq1}
\end{eqnarray}

Here, $H_S$, $H_F$ and $H_{S, F}$ are the von-Neumann entropies of the system, the environmental fragment, and both the system and environmental fragment, respectively.

\section{Experimental realization of Darwinism states in IBM Q devices \label{qd_Sec3}}

\begin{figure}[]
\centering
\includegraphics[scale=0.3]{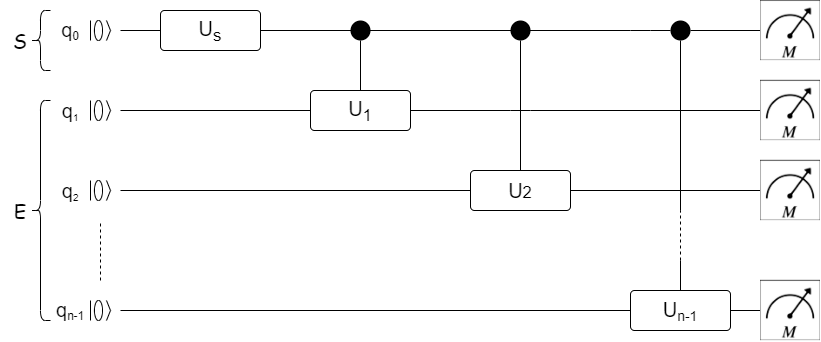}
\caption{The figure shows the quantum circuit consisting n quantum resisters, one single and (n-1) two-qubit quantum gates provided by IBM QX for producing a quantum Darwinism state. We split this n quantum resisters as the first resister representing the system, and the rest (n-1) is for the environment.}
\label{qd_Fig2}
\end{figure}

This section discusses run results of Darwinism state on two different real quantum devices the 5-qubit ``ibmq\_athens" and the 15-qubit ``ibmq\_16\_melbourne" available on the IBM QX open platform. Fig. \ref{qd_Fig2} shows the quantum circuit that we have used in the experiments. The initial state of the circuit is $\ket{00...0_n}$, and after applying the unitary gate ($U_S$) on the first qubit and then a series of control unitary ($cU_i$) operations as shown in the quantum circuit (Fig. \ref{qd_Fig2}), it becomes the quantum Darwinism state given by,

\begin{equation}
\begin{split}
\ket{\psi_N} = \ &\alpha\ket{0}_S \otimes_{i=1}^N \ket{0}_i \\& + \beta\ket{1}_S   \otimes_{i=1}^N
  (\cos{\frac{\theta_i}{2}}\ket{0}_i + \sin{\frac{\theta_i}{2}}\ket{1}_i)
\end{split}
\label{qd_Eq2}
\end{equation}

Here, $\alpha$ and $\beta$ are the normalization constants, and N is the number of environment qubits.

Furthermore, to achieve our tasks (that we discuss further), the QISKit program is quite beneficial because we have to measure many different quantum circuits on different Pauli basis sets. Hence, we write several programs for each experiment that submit jobs to the IBM device as allowed and saves the outcome on our machine for further procedure.

\subsection{Circuit formation}
Here, we use $U_S = U(\theta,0,0)$, $cU_i = cU(\theta_i,0,0,0)$ gate, quantum and classical registers, which are provided on the IBM QX platform \cite{qd_IBMQX}.

\begin{equation}
    U(\theta,\phi,\lambda) = \begin{pmatrix} cos{\frac{\theta}{2}}&-e^{\iota\lambda}sin{\frac{\theta}{2}}\\ e^{\iota\phi}sin{\frac{\theta}{2}}&e^{\iota(\phi+\lambda)}cos{\frac{\theta}{2}} \end{pmatrix} 
\label{qd_Eq3}
\end{equation}

\begin{equation}
    cU(\theta,\phi,\lambda,\gamma) = \begin{pmatrix}
    1&0&0&0\\
    0&1&0&0\\
    0&0&e^{\iota\gamma}cos{\frac{\theta}{2}}&-e^{\iota(\gamma+\lambda)}sin{\frac{\theta}{2}}\\ 0&0&e^{\iota(\gamma+\phi)}sin{\frac{\theta}{2}}&e^{\iota(\gamma+\phi+\lambda)}cos{\frac{\theta}{2}} \end{pmatrix} 
    \label{qd_Eq4}
\end{equation}

\begin{itemize}
    \item First, we create a system's state as $\alpha\ket{0}_S + \beta\ket{1}_S$ (we did not include phase for our experiment) via applying $U_S$ gate on the first $q_0$ qubit. It is to be noted that $\theta =2*\tan^{-1}(\frac{real(\beta)}{\alpha})$. Hence, choosing the correct values of $\theta$, we can generate the desired system's state. For this experiment, we take the value of $\theta$ = $\frac{\pi}{2}$.
    \begin{equation*}
        \ket{00...0_n} = U_s\ket{0}\ket{00...0_n} = \alpha\ket{0}\ket{0...0_n} + \beta\ket{1}\ket{00...0_n} 
    \end{equation*}
    \item After forming the system state, we apply $cU_i$ gate with the system as the control and the targets are the ``environmental" qubits (as shown in Fig. \ref{qd_Fig2}). After these operations, the system entangles with the environment and imprints its information on the environmental qubits. The interaction strength parameters ($\theta_i$) for this $cU_i$ gate are explained in the table \ref{qd_Tab1} for each case with interaction strengths A and B.
    \begin{equation}
        \alpha\ket{00...0_n} + \beta\ket{10...0_n} \xrightarrow[ \textit{keeping system qubits as control}]{\textit{apply cU on environment}} \ket{\psi_N}
        \label{qd_Eq5}
    \end{equation}    
\end{itemize}

\begin{table}[t]
\caption{Interaction strength for each case}
    \centering
    \begin{tabular}{ccc}
    \hline
    \hline
        Case &   Interaction strength  & Interaction strength      \\
           &    A  & B   \\
        \hline
        2-qubit & $\pi$ & $\frac{2\pi}{5}$ \\
        3-qubit & $\pi, \pi$ & $\frac{2\pi}{5}, \frac{5\pi}{9}$ \\
        4-qubit & $\pi, \pi, \pi$ & $\pi, \frac{2\pi}{5}, \frac{5\pi}{9}$ \\
        5-qubit & $\pi, \pi, \pi, \pi$ & $\pi, \pi, \frac{2\pi}{5}, \frac{5\pi}{9}$ \\
        6-qubit & $\pi, \pi, \pi, \pi, \pi$ & $\pi, \pi, \pi, \frac{2\pi}{5}, \frac{5\pi}{9}$ \\
        
        \hline
    \end{tabular}
    \label{qd_Tab1}
\end{table}

\subsection{Quantum State Tomography}
It is a method to reconstruct the state of a quantum system using measurement in different combinations of basis \cite{qd_BeheraQIP2017,qd_VishnuQIP2018,qd_Altepeter2004}. The bases are I, X, Y, Z, known as Pauli bases. It is to be noted that all the measurements are done in Z-basis on the IBM Q platform. Hence, to measure in X and Y bases, we need to apply $H$ and $S^{\dagger}H$ respectively before the Z-basis measurement box.

From this tomography technique, we collect the state in the form of a density matrix, also called the reconstructed experimental density matrix. The expression for the experimental reconstructed density matrix is given as (for n qubit systems),

\begin{equation}
\rho^E =\frac{1}{2^n} \sum_{t_1,t_2,...,t_n=0}^{3} (S_{t_1} * S_{t_2} * ...*S_{t_n})*\sigma_{t_1,t_2,...,t_n}
\label{qd_Eq6}
\end{equation}

where, $\sigma_{t_1,t_2,...t_n}=(\sigma_{t_1}\otimes\sigma_{t_2}\otimes...\otimes\sigma_{t_n})$, and $\sigma_{t_1}$, $\sigma_{t_2}$,..., $\sigma_{t_n}$ represent Pauli matrices. Here, the indices $t_1,t_2,...,t_n$ can take values 0, 1, 2, 3 corresponding to I, X, Y, Z. And, S is the Stokes parameter, which is defined as $S = P_{0} - P_{1}$ for X, Y, Z basis, and $S = P_{0} + P_{1}$ for I basis. Here, we use X, Y, and Z basis for measurement, and did not include the I basis measurement because we can calculate the Stoke parameters value for measurement setting IO and OI from OO measurement settings outcome as \cite{qd_Altepeter2004},

\begin{eqnarray}
S_{OO}=(P0-P1)(P0-P1)=P00-P01-P10+P11\nonumber\\
S_{IO}=(P0+P1)(P0-P1)=P00-P01+P10-P11\nonumber\\
S_{OI}=(P0-P1)(P0+P1)=P00+P01-P10-P11\nonumber\\
\label{qd_Eq7}
\end{eqnarray}

Here, O can be X, Y, and Z basis measurements. Hence, using OO measurement basis probabilities, we can calculate the Stokes parameters, $S_{IO}$ and $S_{OI}$, by changing sign. Thus, in total, we have taken $3^n$ different settings for the n qubit quantum circuit, i.e., 9, 27, 81, 243, and 729 measurement settings are used to perform quantum state tomography of 2-, 3-, 4-, 5-, and 6-qubit quantum circuits respectively.

We reconstruct the experimental density matrices using the real quantum devices, the 15-qubit ibmq\_16\_melbourne, and the 5-qubit ibmq\_athens with 8192 shots for each setting. It is noted that 8192 shots are the maximum number of shots available on the IBM quantum experience platform. A single shot represents a single measurement performed for a particular quantum circuit. The quantum state tomography is also used for calculating the closeness between theoretical and experimental density matrices. This closeness is called fidelity (F), and its expression is given as,

\begin{equation}
    F(\rho^T, \rho^E) =  \textit{Tr}\left(\sqrt{\sqrt{\rho^T}\rho^E\sqrt{\rho^T}}\right)
    \label{qd_Eq8}
\end{equation}

where, $\rho^T = \ket{\psi}\bra{\psi}$ is the theoretical density matrix of the system, which is calculated for N-qubit system as follows, ${\rho^T}_{N}=\ket{\psi_{N}}\bra{\psi_{N}}$.

\begin{table}[t]
\caption{Fidelity results obtained from the real devices ibmq\_16\_melbourne and ibmq\_athens, with and without mitigation with interaction strength parameters A and B. Here, S-$E_N$ represents 1-system qubit and N environmental qubits interaction.}
    \centering
    \begin{tabular}{ccccc}
    \hline
    \hline
        Case &   Athens  & Mitigated & Melbourne & Mitigated      \\
           &      & Athens  &  & Melbourne \\
        \hline
        S-$E_1$ (A) & 0.9427 & 0.9916 & 0.8773 & 0.9855\\
        S-$E_1$ (B) & 0.9568 & 0.9751 & 0.9276 & 0.9663\\
        S-$E_2$ (A) & 0.9143 & 0.9821 & - & -\\
        S-$E_2$ (B) & 0.9261 & 0.9596 & - & -\\
        S-$E_3$ (A) & 0.7419 & 0.7933 & 0.6069 & 0.9681\\
        S-$E_3$ (B) & 0.7603 & 0.7412 & 0.5667 & 0.7658\\
        S-$E_4$ (A) & 0.7260 & 0.8101 & 0.2210 & 0.3533\\
        S-$E_4$ (B) & 0.7260 & 0.7915 & 0.1318 & 0.1848\\
        S-$E_5$ (A) & - & - & 0.0557 & 0.0990\\
        S-$E_5$ (B) & - & - & 0.0727 & 0.0896\\
        \hline
    \end{tabular}
    \label{qd_Tab2}
\end{table}

\begin{table}[t]
\caption{Purity of the density matrix obtained from the real devices ibmq\_16\_melbourne and ibmq\_athens, with and without mitigation with interaction strength parameters A and B. Here, S-$E_N$ represents 1-system qubit and N environmental qubits interaction.}
    \centering
    \begin{tabular}{ccccc}
    \hline
    \hline
        Case &   Athens  & Mitigated & Melbourne & Mitigated      \\
           &      & Athens  &  & Melbourne \\
        \hline
        S-$E_1$ (A) & 0.8924 & 0.9854 & 0.7861 & 0.9852\\
        S-$E_1$ (B) & 0.9236 & 0.9630 & 0.8696 & 0.9458\\
        S-$E_2$ (A) & 0.8479 & 0.9747 & - & -\\
        S-$E_2$ (B) & 0.8736 & 0.9394 & - & -\\
        S-$E_3$ (A) & 0.6140 & 0.7167 & 0.4319 & 1\\
        S-$E_3$ (B) & 0.6891 & 1 & 0.4736 & 1.2\\
        S-$E_4$ (A) & 0.6086 & 0.7518 & 0.1988 & 0.5622\\
        S-$E_4$ (B) & 0.6126 & 0.7226 & 0.2253 & 0.5242\\
        S-$E_5$ (A) & - & - & 0.1346 & 0.4035\\
        S-$E_5$ (B) & - & - & 0.1294 & 0.3370\\
        \hline
    \end{tabular}
    \label{qd_Tab3}
\end{table}

In Table \ref{qd_Tab2} and Table \ref{qd_Tab3} we have presented the fidelities and purity of $\rho^E$ (with and without mitigation) for 2-, 3-, 4-, 5- and 6- qubit systems after executing the quantum circuits. It can be observed that in each of the cases, after performing the mitigation, the fidelity and purity both are increased (it means that after performing the mitigation process, the obtained density matrix is more close and pure than without mitigation). Also, the fidelities and purity of the density matrix obtained from ibmq\_athens are much higher from ibmq\_16\_melbourne for the five qubit system, and for the rest, it is a bit higher. However, overall, there is a decrease in the state's fidelity and purity as we go from 2- to 6-qubits system for both devices. The main reason for the above is that as we increase the number of qubits in the quantum circuits, the number of measurement settings exponentially increases in the order of $3^n$. We know that each execution result contains some error, consequently accommodating an exponential increase in error to perform quantum state tomography. Athens seems to have relatively lower gate errors than the Melbourne one, depending upon the real chip performance that includes the characterization of device parameters such as qubit alignments, the number of qubits, coherence time, and qubit errors gate errors, to name a few. Furthermore, in the higher number qubit system, we implement more two-qubit control gates; if they are decomposed, it introduces more single-qubit and two-qubit gates, resulting in more gate errors. Moreover, we have transpiled all our quantum circuits with a higher level of optimization in each measurement setting before executing on the real devices to obtain the experimental states with maximum fidelity.

\section{Mutual, Quantum and Classical Information \label{qd_Sec4}}

\begin{figure*}
\includegraphics[width=0.9\textwidth]{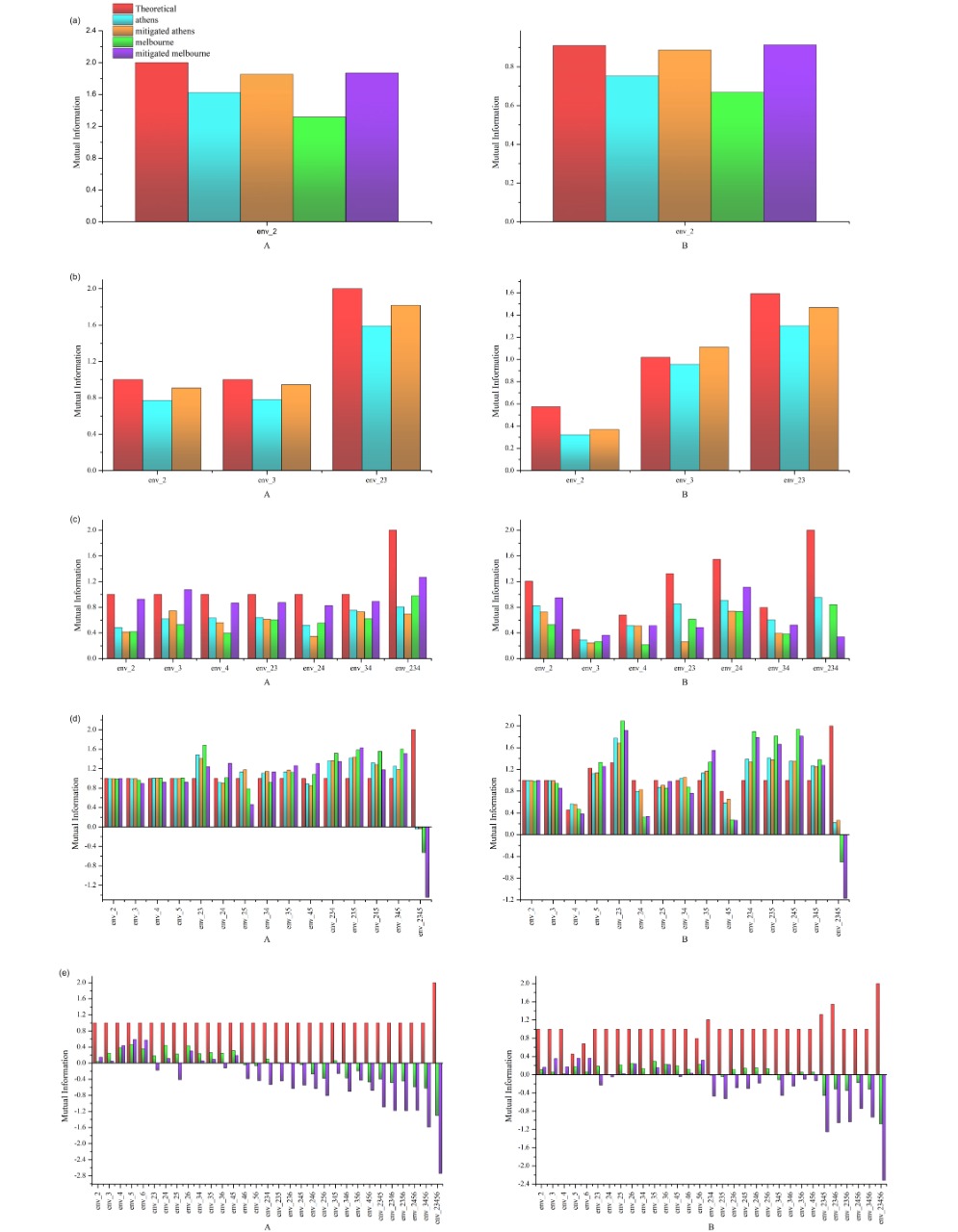}
\caption{Mutual information (MI) plots between the system and different environmental fragments for the 1-, 2-, 3-, 4-, and 5-qubit as the environment are given in (a), (b), (c), (d), and (e) respectively. On the left and right columns, we show the results for interaction strength A and B, respectively. In red, we show theoretical; in cyan and orange, we show Athens with and without mitigation, respectively; In green and violet, we show Melbourne with and without mitigation, respectively MI results.}
\label{qd_Fig3}
\end{figure*}

\subsection{Mutual Information}
From \ref{qd_Eq1}, the mutual information between system and environmental fragments is defined as 
\begin{equation}
    I(S:E_i) = H(\rho_{S}) + H(\rho_{E_i}) - H(\rho_{SE_i})
    \label{qd_Eq9}
\end{equation}

Here, $\rho_{S}$ and $\rho_{E_i}$ are the reduced density matrices for system and environmental fragments, respectively, and $H$ is the Von-Neumann entropy. This equation \ref{qd_Eq9} quantifies the level of information present in the environment about the system. When this MI value is 1, $E_i$ has the complete record of the system $S$, and if it is lower than 1, it does not have the complete record of $S$. If it exceeds the value, then it has two cases; first, $E_i$ is the whole environment, and secondly, when it is a fragment of the whole environment, it has more information than the system $S$.

In our experiment, we successfully calculate mutual information for $\rho^T$, $\rho^E$, and mitigated $\rho^E$, shown in Fig. \ref{qd_Fig3}. We perform error mitigation on experimental results to increase their accuracy. From the figure, we notice that the MI results for $\rho^T$ are identical according to the theory \cite{qd_Zureknature2009}. However, the real device gives satisfactory results for the lower number of qubit systems (2- and 3-qubit systems). Furthermore, as we increase the number of environmental qubits (for 4-, 5- and 6-qubit systems), MI's value becomes unsatisfactory or unexplainable. From the MI theory, it is clear that the value of MI is always positive \cite{qd_Wittenspringer2020}, but here we obtain some negative values too. These odd results are due to the noise and error introduced during calculating the density matrices, as discussed in the quantum state tomography part. Also, it can be mentioned that while interaction angle is $\frac{2\pi}{5}$ or $\frac{2\pi}{5}$, the obtained information is lesser than when it is $\pi$. Because, with $\pi$, the environmental qubit is highly correlated with the systems qubit. 

\subsection{Holevo quantity} The Holevo quantity $\chi$ is the measurement of the classical information passed by the quantum channel (here environment $E$) about the observable (here system $S$), and this quantity is calculated as

\begin{equation}
    \chi(S,E_i) = H(\sum_{s}p_s\rho_{E_{i}|s}) - \sum_sp_sH(\rho_{E_{i}|s})
    \label{qd_Eq10}
\end{equation}

Here, $\rho_{E_{i}|s}$ represents the density matrix of the environmental fragment conditioned on outcome of the system's measurement and $p_s$ is the outcome probability of system while doing the measurement. Here, we calculate this value as,

\begin{equation}
    \ket{\psi_N} = \alpha\ket{0}_S \otimes_{i=1}^N \ket{0}_i + \beta\ket{1}_S \otimes_{i=1}^N (\cos{\frac{\theta_i}{2}}\ket{0}_i + \sin{\frac{\theta_i}{2}}\ket{1}_i)
    \label{qd_Eq11}
\end{equation}

Here, the first qubit shows the system and the rest are for the environment, and the state becomes as follows,

\begin{equation}
    \alpha\ket{0}_s\ket{\psi_{E_{i}|0}} + \beta\ket{1}_s\ket{\psi_{E_{i}|1}} 
    \label{qd_Eq12}
\end{equation}
and the density matrix is $\rho_{E_{i}|0} = \ket{\psi_{E_{i}|0}}\bra{\psi_{E_{i}|0}}$ and $\rho_{E_{i}|1}=\ket{\psi_{E_{i}|1}}\bra{\psi_{E_{i}|1}}$.

\subsection{Discord}
It is the measurement of quantum information present in the environment about the system and defines as

\begin{equation}
    D(S,E_i) = I(S,E_i) - \chi(S,E_i)
    \label{qd_Eq13}
\end{equation}

\begin{figure}
\includegraphics[width=8cm]{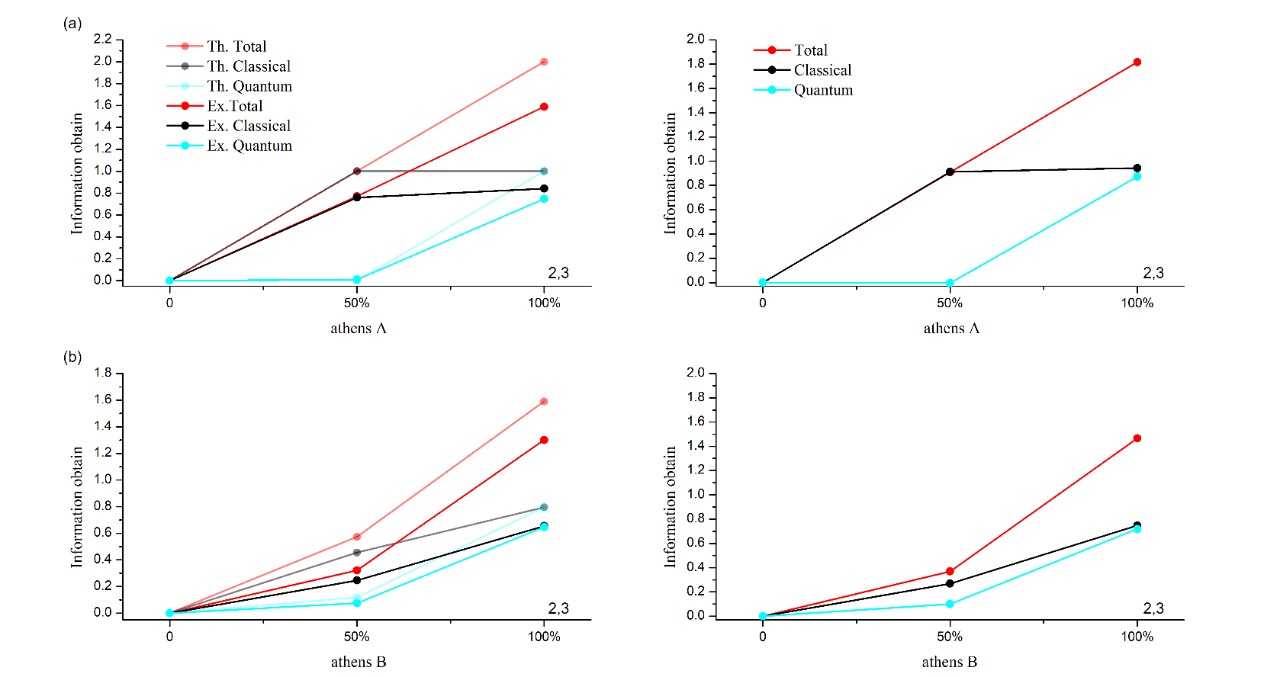}
\caption{Comparison of obtained information for 1 qubit system and 2 qubits environment. (a) Quantum Darwinism process with interaction strength A. (b) Quantum Darwinism process with interaction strength B.}
\label{qd_Fig5}
\end{figure}

\begin{figure}[]
\includegraphics[width=8cm]{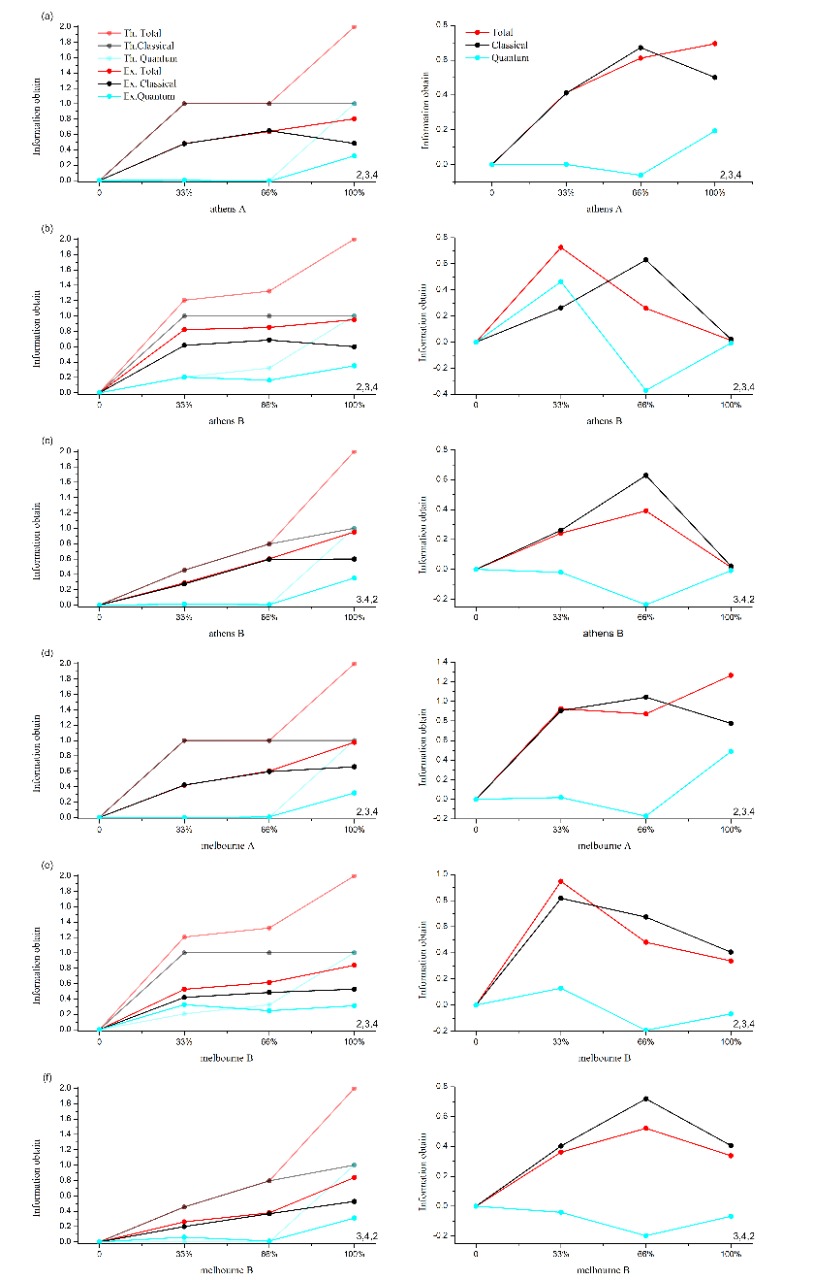}
\caption{Comparison of obtained information for 1 qubit system and 3 qubits environment. (a),(d) Quantum Darwinism process with interaction strength A. (b),(c),(e),(f) Quantum Darwinism process with interaction strength B.}
\label{qd_Fig6}
\end{figure}

\begin{figure}[]
\includegraphics[width=8cm]{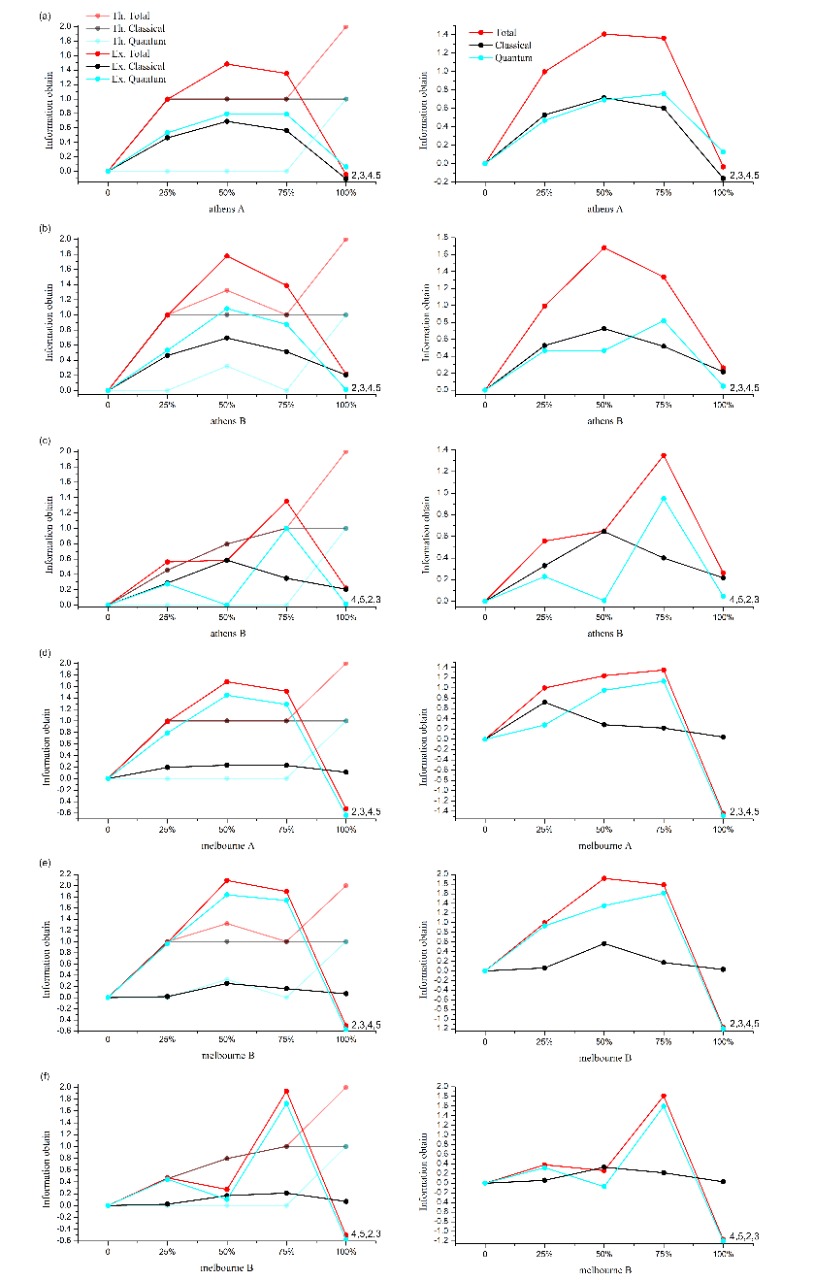}
\caption{Comparison of obtained information for 1 qubit system and 4 qubits environment. (a),(d) Quantum Darwinism process with interaction strength A. (b),(c),(e),(f) Quantum Darwinism process with interaction strength B.}
\label{qd_Fig7}
\end{figure}

\begin{figure}
\includegraphics[width=8cm]{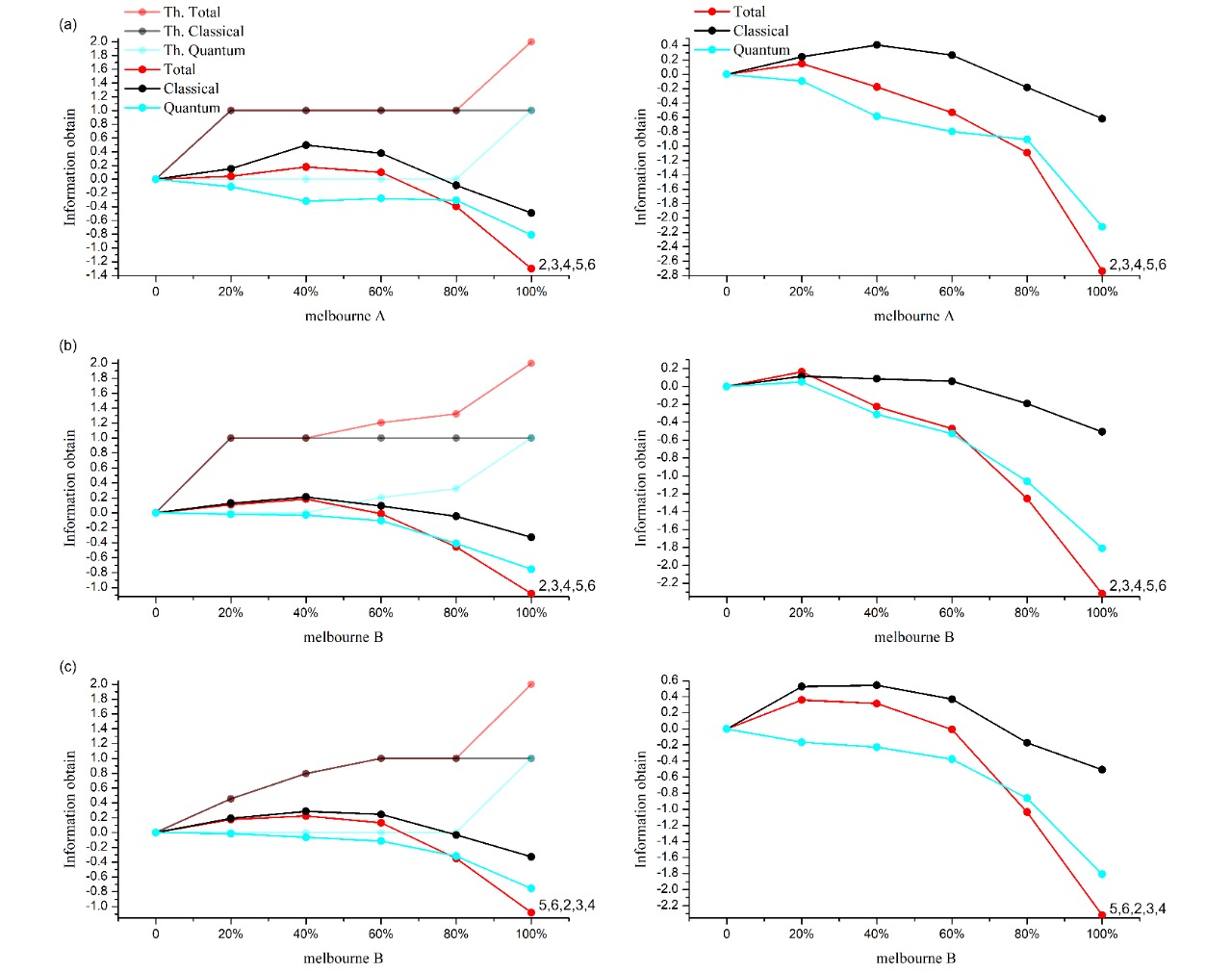}
\caption{Comparison of obtained information for 1 qubit system and 5 qubits environment. (a) Quantum Darwinism process with interaction strength A. (b),(c) Quantum Darwinism process with interaction strength B.}
\label{qd_Fig8}
\end{figure}

The discord (D) and Holevo ($\chi$) plots are correlated \cite{qd_ZwolakSciRep2013}, such as if the Holevo plot rises to the classical plateau early as we select the minimal fragment of the total environment then discord only rises when we select the whole environment. As we observe in the interaction strength A case (Figs. \ref{qd_Fig5} (a), \ref{qd_Fig6} (a,d), \ref{qd_Fig7} (a,d), \ref{qd_Fig8} (a) see light black and cyan line), and if we see delayed rise in the Holevo quantity toward the plateau then consecutively discord starts rising earlier as in the interaction strength B case (Figs. \ref{qd_Fig5} (b), \ref{qd_Fig6} (b,c,e,f), \ref{qd_Fig7} (b,c,e,f), \ref{qd_Fig8} (b,c) see light black and cyan line.

\begin{figure}
\includegraphics[width=8cm]{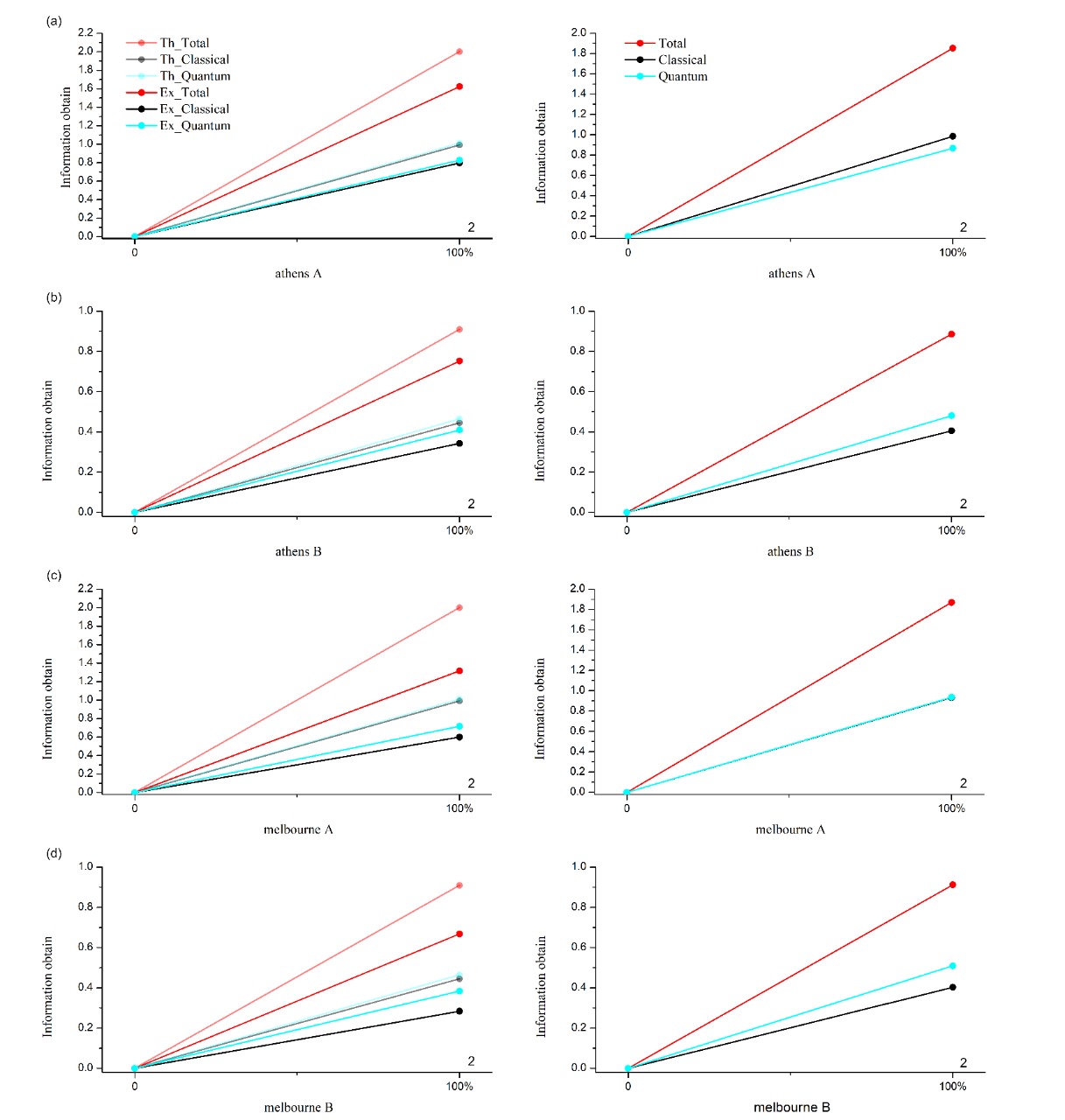}
\caption{Comparison of obtained information for 1 qubit system and 1 qubit environment. (a),(c) Quantum Darwinism process with interaction strength A. (b),(d) Quantum Darwinism process with interaction strength B.}
\label{qd_Fig4}
\end{figure}

\subsection{Discussion on obtained clasical-quantum information}
The figures (\ref{qd_Fig4}, \ref{qd_Fig5}, \ref{qd_Fig6}, \ref{qd_Fig7}, \ref{qd_Fig8}) show the differences in information obtained from theoretical, device and mitigated computation. The left column shows theoretical and experimental results, whereas the right column shows the mitigated experimental results. Here, light red, black, and cyan lines represent theoretical total, classical, and quantum correlations. In contrast, the dark red, black, and cyan lines represent experimental/mitigated total, classical, and quantum correlations between the system and the environment. X and Y axis show fraction of environments and obtained information respectively, the X-axis label is divided such that \{device name\} \{interaction strength\} (Ex:- Athens B). The theoretical results are identical with the quantum-classical correlation theory \cite{qd_ZurekAP2000, qd_OllivierPRL2002, qd_HendersonJPA2001} (statement from the theory is that the classical information can be achieved by accessing some fragments of the environment whereas the quantum information can be observed only when the fragment is the whole environment). However, the rest of the results are not identical to the theoretical because in the experimental case, we obtain a mixed state instead of a pure as in the theoretical case, and the level of fidelity and purity we have shown in the Table \ref{qd_Tab2} and Table \ref{qd_Tab3} respectively. This mixed and in-close states result variation of information obtained in each case as we observe from mutual information \ref{qd_Fig3}, and classical-quantum information \ref{qd_Fig4}, \ref{qd_Fig5}, \ref{qd_Fig6}, \ref{qd_Fig7}, \ref{qd_Fig8} between system and different environmental fragments. Moreover, it is to be noted that the higher the state is mixed the higher its von-Neumann entropy (for 100\% pure system it is zero) means highly mixed state contain more missing information about the system. Here, the order of environment in figures \ref{qd_Fig4}, \ref{qd_Fig5}, \ref{qd_Fig6}, \ref{qd_Fig7}, \ref{qd_Fig8} from 0 to 100\% follows the order as written in the plot on X-axis. Ex:- if the order is (a,b,c,d) the 25\% is a, 50\% is (a, b), 75\% is (a, b, c), and 100\% is (a, b, c, d).

\section{Discussions and Conclusion \label{qnm_Sec4}}
To conclude, we have created a quantum circuit \ref{qd_Fig2} using n quantum registers for the environmental interaction model \ref{qd_Fig1} and using the quantum circuit, the quantum Darwinism state (Eq. \eqref{qd_Eq2}) has been successfully constructed for n = 2, 3, 4, 5 and 6. The data has been collected from real quantum devices \textbf{ibmq\_athens} and \textbf{ibmq\_16\_melbourne} (every execution has 8192 shots) using quantum state tomography, which is in form of density matrix. Using this density matrix, we have calculated the fidelity and purity to check the quality of experimental results, and the outcomes have been arranged in table \ref{qd_Tab2} and \ref{qd_Tab3} respectively. 

The mutual information between the system and the environment has been calculated to observe how much record the environment has for the system state. Also, the classical and quantum information of the system delivered by the environmental qubit is quantified using the Holevo and quantum discord. We have plotted this information results and shown them in figures \ref{qd_Fig3}, and  \ref{qd_Fig4}, \ref{qd_Fig5}, \ref{qd_Fig6}, \ref{qd_Fig7}, \ref{qd_Fig8} respectively. Here, the Holevo quantity shows that to access the classical information about the systems state, and we do not need to measure the whole environment; just a tiny fragment is sufficient. However, we have to measure the total environment for quantum information because it can not be shared between the environment fragments. Both conclusions are according to the theoretical results. From experimental results, this explanation holds for n = 2 and 3. However, for the rest, the introduced noise and error during execution, accommodates exponential increase in error as the n-value increases further. In achieving all these tasks, we write a QISKit program for every case because it is an easy way to submit multiple jobs on IBM QX and perform further computation using the run results.

\section*{Acknowledgments}
R.S. would like to thank Bikash's Quantum (OPC) Pvt. Ltd. for providing hospitality during the course of the project work. The authors acknowledge the support of IBM Quantum Experience. The views expressed are those of the authors and do not reflect the official policy or position of IBM or the IBM Experience team.

\end{document}